# Machine Parameters and Projected Luminosity Performance of Proposed Future Colliders at CERN

*F. Bordry, M. Benedikt, O. Brüning, J. Jowett, L. Rossi, D. Schulte, S. Stapnes, F. Zimmermann,*
CERN, Geneva, Switzerland

**Abstract**

In response to a request from the CERN Scientific Policy Committee (SPC), the machine parameters and expected luminosity performance for several proposed post-LHC collider projects at CERN are compiled: three types of hadron colliders (HL-LHC upgrade, FCC-hh and HE-LHC), a circular lepton collider (FCC-ee), a linear lepton collider (CLIC), and three options for lepton-hadron colliders (LHeC, HE-LHeC, and FCC-eh). Particular emphasis is put on availability, physics run time, and efficiency. The information contained in this document was presented at the SPC Meeting of September 2018. It will serve as one of the inputs to the 2019/20 Update of the European Strategy for Particle Physics.

# Contents

# 1    Introduction

This document discusses the main machine parameters and expected luminosity performance for several proposed post-LHC collider projects at CERN: three types of hadron colliders (the HL-LHC upgrade, FCC-hh and HE-LHC), a circular lepton collider (FCC-ee), a linear lepton collider (CLIC) and three options for lepton-hadron colliders (LHeC, HE-LHeC and FCC-eh).

# 2    Circular Proton Colliders: HL-LHC, FCC-hh and HE-LHC

## 2.1    HL-LHC

The High-Luminosity LHC (HL-LHC) [1] is an approved luminosity upgrade of the LHC [2]. The HL-LHC will be the first accelerator ever to feature multiple quadrupole and dipole magnets based on $Nb_3Sn$ superconductors, thereby constituting an important technological stepping stone towards longer-term future projects like FCC-hh and HE-LHC. The HL-LHC will also be the first hadron collider to use crab cavities. The peak luminosity for the high-luminosity experiments ATLAS and CMS will be levelled at a constant value of around $5 \times 10^{34}$ cm$^{-2}$s$^{-1}$ ($7.5 \times 10^{34}$ cm$^{-2}$s$^{-1}$ ultimate value), in order to limit radiation and event pile-up in the experiments. Expected to operate from 2026 through to the late 2030s, the HL-LHC will increase the integrated luminosity of the LHC by an order of magnitude, yielding a total accumulated value above 3 ab$^{-1}$ (4 ab$^{-1}$ ultimate value).

The upgraded heavy-ion programme at LHC [3], starting immediately after LS2, will deliver nucleus-nucleus (Pb-Pb) and proton-nucleus (p-Pb) luminosities several times beyond design. After a successful Xe-Xe pilot run in 2017 [4], collisions of lighter nuclei are also under consideration by the physics community because they may provide significantly higher nucleon-nucleon luminosity.

## 2.2    FCC-hh

The FCC-hh [5, 6] would provide proton-proton collisions with a centre-of-mass energy of 100 TeV, about seven times higher than the LHC, and be able to accumulate an integrated luminosity of ~ 20 ab$^{-1}$ in 25 years of operation. With the initial parameters, a maximum luminosity of $5 \times 10^{34}$ cm$^{-2}$s$^{-1}$ is expected to be reached in the first few years. In the following fifteen years, with the nominal parameters, the luminosity could reach up to $3 \times 10^{35}$ cm$^{-2}$s$^{-1}$. The FCC-hh nominal bunch population is $10^{11}$ and the normalised transverse rms emittance at the start of a store is 2.5 μm. The current LHC, as of early autumn 2018, operates with $1.1 \times 10^{11}$ particles per bunch and an initial normalised transverse rms emittance of 2.0 μm.

The FCC-hh also offers the potential to collide ions with protons and ions with ions. In addition, the design allows for an upgrade to electron-proton and electron-ion collisions. In these cases, one of the circulating proton or ion beams would collide with an electron beam provided by a new energy recovery linac.

## 2.3    HE-LHC

Reaching a target beam energy around 13.5 TeV in the existing LHC tunnel would rely on the FCC-hh magnet technology. The present LHC dipole magnets, which have a nominal field of 8.33 Tesla, would be replaced by FCC-type 16 Tesla dipole magnets. Achieving a centre-of-mass energy close to 27 TeV with 16 T magnets requires a dipole-filling factor similar to that of the LHC.

An integrated luminosity exceeding 10 ab$^{-1}$ is within reach over about 20 years of operation of the HE-LHC [7]. After the LHC Injector Upgrade (LIU) [8], scheduled for 2020, an extremely bright proton beam will be available for injection into the HE-LHC. Assuming essentially the same beam parameters as for the HL-LHC, the HE-LHC bunch population is taken to be $2.2 \times 10^{11}$ and the normalised transverse rms emittance at the start of a store to be 2.5 µm. Bunch spacing of 25 ns has been chosen to be the same as in the LHC, HL-LHC and FCC-hh. The luminosity evolution during HE-LHC physics is determined by the combined effects of proton burn-off and significant radiation damping.

In addition to delivering proton-proton physics at the energy frontier, the HE-LHC could operate as a heavy-ion and ion-proton collider.

By adding a 60 GeV electron beam from a multi-pass energy-recovery linac, the HE-LHC could also provide high-energy lepton-proton and lepton-ion collisions (HE-LHeC).

## 2.4  Proton-Proton Collisions

The baseline design parameters for HE-LHC and FCC-hh are summarised in Table 1, which also presents a comparison with the corresponding values for LHC and HL-LHC. It is assumed that both the FCC-hh and the HE-LHC will accommodate two high-luminosity interaction-points (IPs) and that two other IPs could host secondary experiments.

The machine availability $A$ for all hadron colliders, including their injectors, is assumed to be at least 75%, and to reach 80% for the HL-LHC (similar to the 2017 performance of the LHC).

The fraction of time in physics, $f_{phys}$ is estimated through the optimum physics run time (without failure) $t_{run}$, the average turnaround time $t_{ta}$ and the availability as [5]:

$$f_{\text{phys}} \approx A \frac{t_{run}}{t_{run}+t_{ta}}$$

Similarly, the effect of the availability $A$ on the integrated luminosity after a time T in days is approximated as:

$$L_{\text{int}} \approx A \, T \, L_{\text{nominal per day}}$$

where $L_{\text{nominal per day}}$ denotes the nominal integrated luminosity per day, for acycle without any hardware failure. Analytical expressions for $L_{\text{nominal per day}}$ depend on the relative strength of radiation damping and proton burn-off [5, 6], and on whether or not a levelling scheme is applied [5].

## 2.1  Heavy-Ion Operation

To support a second major physics programme, the FCC-hh or HE-LHC machines could operate as ion-ion or proton-nucleus colliders, similar to the successful heavy-ion programme at the LHC [3].

Considering the current baseline assumptions for the proton-proton programme, presently the main focus lies on collisions of lead nuclei with each other (Pb-Pb) and with protons (p-Pb).

Beam and luminosity evolution, as well as estimates for the integrated luminosity, are available for heavy-ion operation at FCC-hh. The FCC-hh performance projections assume the LHC to be the final injector synchrotron. In order to provide a broad overview of the potential as an ion collider, several operational scenarios are compared:

1. Two beam parameter cases (see Table 2): *initial and nominal*, which differ in the β-function at the interaction point (IP) and the bunch spacing, which defines the maximum number of circulating bunches.
2. Collisions at one or two experiments, which share the total luminosity available.

**Table 1:** Key parameters of HE-LHC and FCC-hh compared with HL-LHC and LHC, for operation with proton beams. All values, except for the injection energy, refer to collision energy.

| Parameter | Unit | FCC-hh | | HE-LHC | (HL-)LHC |
|---|---|---|---|---|---|
| Centre-of-mass energy | TeV | 100 | | 27 | 14 |
| Injection energy | TeV | 3.3 | | 1.3 (0.9) | 0.45 |
| Arc dipole field | T | 16 | | 16 | 8.33 |
| Circumference | km | 97.8 | | 26.7 | 26.7 |
| Straight-section length | m | 1400 | | 528 | 528 |
| Beam current | A | 0.5 | | 1.12 | (1.12) 0.58 |
| Bunch population | $10^{11}$ | 1.0 | | 2.2 | (2.2) 1.15 |
| Number of bunches / beam | — | 10400 | | 2808 | (2760) 2808 |
| RF voltage | MV | 24 | | 16 | (16) 16 |
| RMS bunch length | mm | $\sim 80$ | | 90 | (90) 75.5 |
| Bucket half height | $10^{-3}$ | 0.075 | | 0.21 | 0.36 |
| RMS momentum spread | $10^{-4}$ | 0.3 | | 0.85 | 1.129 |
| Longit. emittance ($4\pi\sigma_z\sigma_E$) | eVs | 5 | | 4.2 | 2.5 |
| Bunch spacing | ns | 25 | | 25 | 25 |
| Norm. transv. rms emittance | $\mu$m | 2.2 | | 2.5 | (2.5) 3.75 |
| IP beta function $\beta^*_{x,y}$ | m | 1.1 | 0.3 | 0.45 | (0.15) 0.55 |
| Initial rms IP beam size $\sigma^*_{x,y}$ | $\mu$m | 6.7 | 3.5 | 9.0 | (7.1 min.) 16.7 |
| Half crossing angle | $\mu$rad | 37 | 100 | 165 | (250) 142.5 |
| Piwinski angle w/o crab cav. | — | 0.42 | 2.16 | 1.7 | (2.7) 0.65 |
| Peak luminosity per IP | $10^{34}$ cm$^{-2}$s$^{-1}$ | 5 | 30 | 16 | (5, levelled) 1 |
| Total cross section | mbarn | 153 | | 126 | 111 |
| Inelastic cross section | mbarn | 108 | | 91 | 85 |
| Peak no. of events / crossing | — | 170 | 1000 | 460 | (135) 27 |
| RMS luminous region | mm | 53 | 49 | 57 | (40) 45 |
| Stored energy / beam | GJ | 8.4 | | 1.4 | (0.7) 0.36 |
| Energy loss / proton / turn | keV | 4600 | | 93 | 6.7 |
| SR power / beam | kW | 2400 | | 100 | (7.3) 3.6 |
| SR power / length | W/m/ap. | 29 | | 4.6 | (0.33) 0.17 |
| Transv. emit. damping time | h | 1.1 | | 3.6 | 25.8 |
| No. of high-luminosity IPs | — | 2 | 2 | 2 | (2) 2 |
| Initial proton burn-off time | h | 17 | 3.4 | 2.5 | (15) 40 |
| Allocated physics time / year | days | 160 | 160 | 160 | (160) 160 |
| Average turnaround time | h | 5 | 4 | 5 | (4) 5 |
| Optimum run time | h | 11.6 | 3.7 | 5.3 | (18–13) $\sim$10 |
| Accelerator availability | — | 70% | 70% | 75% | (80%) 78% |
| Nominal luminosity per day | fb$^{-1}$ | 2.0 | 8.0 | 5.0 | (1.9) 0.4 |
| Luminosity / year (160 days) | fb$^{-1}$ | $\geq$ 250 | $\geq$ 1000 | 500 | (350) 55 |

Heavy-ion operation is also possible at the HE-LHC. It is assumed that the HE-LHC will use the same Pb beams as the HL-LHC, in terms of charge and normalised emittance. For only two experiments taking luminosity (rather than three or four at HL-LHC), at roughly twice the energy, one can estimate that the Pb-Pb integrated luminosity per experiment would be about a factor of 2 to 3 better than for HL-LHC because of the higher energy, fast radiation damping and shorter fills. The expected integrated luminosity then amounts to around 10 nb$^{-1}$ per one-month run. The peak luminosity could be well over $2 \times 10^{30}$ cm$^{-2}$s$^{-1}$ but would probably be levelled to something lower.

For p-Pb collisions, a similar rough scaling would lead to integrated luminosities of the order of 2 pb$^{-1}$ per one-month run and a peak luminosity of around $2 \times 10^{30}$ cm$^{-2}$s$^{-1}$. These values fall between those for HL-LHC and FCC-hh.

Lighter nuclei are being considered as an option for HL-LHC because they may provide significantly higher nucleon-nucleon luminosity. At HE-LHC and FCC-hh, they may also be preferred as a means to reduce beam losses that limit luminosity.

**Table 2:** Beam and machine parameters for FCC-hh heavy-ion operation

|  | Unit | Baseline | | Nominal | |
|---|---|---|---|---|---|
|  |  | Pb-Pb | p-Pb$^a$ | Pb-Pb | p-Pb$^a$ |
| Operation mode | - | Pb-Pb | p-Pb$^a$ | Pb-Pb | p-Pb$^a$ |
| Beam energy | TeV | 4100 | 50 | 4100 | 50 |
| Centre-of-mass energy / nucleon | TeV | 39.4 | 62.8 | 39.4 | 62.8 |
| No. of bunches | - | 2760 | | 5400 | |
| Bunch spacing | ns | 100 | | 50 | |
| No. of particles / bunch | $10^8$ | 2 | 164 | 2 | 164 |
| Transv. norm. emittance | $\mu$m.rad | 1.5$^b$ | 3.75$^b$ | 1.5$^b$ | 3.75$^b$ |
| Stored energy / beam | MJ | 362 | | 709 | |
| Stored energy / beam at inj. | MJ | 24 | | 47 | |
| $\beta$-function at the IP | [m] | 1.1 | | 0.3 | |
| Number of IPs in collision | - | 1 or 2 | | 1 or 2 | |
| Initial luminosity | $10^{27}$ cm$^{-2}$s$^{-1}$ | 34 | 2800 | 248 | 20400 |
| Peak luminosity$^c$ | $10^{27}$ cm$^{-2}$s$^{-1}$ | 80 | 13300 | 320 | 55500 |
| Integrated luminosity$^d$ (1 exp.) | nb$^{-1}$/run | 35 | 8000 | 110 | 29000 |
| Integrated luminosity$^d$ (2 exp.) | nb$^{-1}$/run | 23 | 6000 | 65 | 18000 |
| Total cross-section | b | 597 | 2 | 597 | 2 |
| Peak BFPP beam power | kW | 19 | 0 | 75 | 0 |

$^a$ p–Pb operation uses the same Pb beam as in Pb-Pb operation. The parameters listed in this column correspond to the proton beam.

$^b$ Pb emittances are based on LHC experience. Proton emittances are chosen to feature the same geometric beam size, which is larger than in p-p operation, where $\epsilon_n = 2.2\,\mu$m.

$^c$ One experiment in collisions.

$^d$ Per experiment. Including an efficiency factor of 50% to take account of down time.

## 2.2 Hadron Collider Performance Assumptions

For all three hadron colliders (HL-LHC, HE-LHC and FCC-hh), the assumption is 160 days scheduled for proton physics per year, as is approximately the case for the present LHC.

This number is derived by subtracting from one year (365 days):

- 125 days(18 weeks) for an end-of-year technical stop (YETS),
- 40 days for commissioning (powering tests and beam commissioning),
- 20 days for machine development (MD),
- 5 days for special runs,
- 10 days for technical stops and another 5 days for recovery after a technical stop.

Table 3 compares the assumed availability values and resulting fractions of time in physics ("stable beams", based on the nominal operation cycle) for HL-LHC, HE-LHC and FCC-hh with the 2017 performance of the LHC.

**Table 3:** Machine availability and fractions of time in physics for HL-LHC, HE-LHC and FCC-hh with actual data from the LHC in 2017.

| Machine | Availability [%] | Fraction of Time in Physics [%] |
|---|---|---|
| LHC 2017 *achieved* | 79 | 50 |
| HL-LHC *assumed* | 80 | 55 |
| HE-LHC *assumed* | 75 | 40 |
| FCC-hh (initial) *assumed* | 70 | 50 |
| FCC-hh (nominal) *assumed* | 70 | 35 |

## 3    Circular Lepton Collider: FCC-ee

### 3.1    Operation Modes and Parameters

The baseline FCC-ee features four modes of operation: (1) on the Z pole, (2) at the WW threshold, (3) at the HZ production peak, and (4) at the ttbar threshold. The four running modes define a 14-year life-cycle for the collider , and are separated by RF system re-configuration periods. Running modes (1) to (3) are combined into Phase 1. Running mode (4) requires a major reconfiguration and is therefore called Phase 2 (365 GeV c.m.).

The physics goals of FCC-ee require the following integrated luminosities for the different operation modes, summed over two interaction points (IPs) [9, 10]:

-    150 ab$^{-1}$ at and around the Z pole (88, 91, 94 GeV centre-of-mass energy),
-    10 ab$^{-1}$ at the W± threshold (~161 GeV with ± a few GeV scan),
-    5 ab$^{-1}$ at the ZH maximum (~ 240 GeV),
-    1.5 ab$^{-1}$ at and above the ttbar threshold (a few hundred fb$^{-1}$) with a scan from 340 to 350 GeV, and the remainder at 365 GeV.

FCC-ee machine parameters for all modes of operation are summarised in Table 4.

### 3.2    Estimating Annual Performance

The annual luminosity estimates for FCC-ee at each mode of operation are derived from three parameters:

-    Nominal luminosity *L* taken to be 10-15% lower than the luminosity simulated for the baseline beam parameters. This nominal luminosity is considered from the third year onward in Phase 1 (Z pole), and from the second year in Phase 2 (ttbar threshold).
-    The luminosity for the first and second year of phase 1 and for the first year of phase 2 are assumed to be smaller, on average, by another factor or two, in order to account for the initial operation.
-    It is assumed that 185 days per year are scheduled for physics. This number of 185 days is obtained by subtracting from one year (365 days):
     -    120 days (17 weeks) for an end-of-year technical stop  (YETS),
     -    30 days of annual commissioning,
     -    20 days for machine development,
     -    10 days for technical stops.
-    Nominal luminosity *L* and time for physics *T* are converted into integrated luminosity Lint via an efficiency factor *E*, according to: $L_{int} = E \cdot T \cdot L$

The efficiency factor *E* is an empirical factor, whose value can be extrapolated from other similar machines, or by simulations with average failure rate and average downtime. Thanks to the top-up mode of operation, it is expected that $E$ will be, within five percent, equal to the availability of the collider complex. We assume an availability of at least 80% and, thereby, a corresponding efficiency $E > 75\%$.

In the case of FCC-ee, no time is lost for acceleration and the efficiency only reflects the relative downtime due to technical problems and associated re-filling and recovery time. Therefore, the efficiency will be roughly equal to the hardware availability, taken to be at least 80%, minus 5% reduction for beam recovery after a failure. The assumed efficiency value of 75% with respect to the daily peak luminosity is lower than achieved with top-up injection at KEKB and PEP-II.

**Table 4:** Key parameters of the FCC-ee circular collider (SR: synchrotron radiation; BS: beamstrahlung)

| | **Z** | W$^\pm$ | **ZH** | **t̄t** |
|---|---|---|---|---|
| Circumference [km] | 97.76 | 97.76 | 97.76 | 97.76 |
| Bending radius [km] | 10.76 | 10.76 | 10.76 | 10.76 |
| Free length to IP $l^*$ [m] | 2.2 | 2.2 | 2.2 | 2.2 |
| SR power / beam [MW] | 50 | 50 | 50 | 50 |
| Beam energy [GeV] | 45.6 | 80 | 120 | 182.5 |
| Beam current [mA] | 1390 | 147 | 29 | 5.4 |
| Bunches / beam | 16640 | 2000 | 328 | 48 |
| Bunch population [$10^{11}$] | 1.7 | 1.5 | 1.8 | 2.3 |
| Horizontal emittance $\varepsilon_x$ [nm] | 0.27 | 0.84 | 0.63 | 1.46 |
| Vertical emittance $\varepsilon_y$ [pm] | 1.0 | 1.7 | 1.3 | 2.9 |
| Arc cell phase advances [deg] | 60/60 | 60/60 | 90/90 | 90/90 |
| Momentum compaction $\alpha_p$ [$10^{-6}$] | 14.8 | 14.8 | 7.3 | 7.3 |
| Horizontal $\beta_x^*$ [m] | 0.15 | 0.2 | 0.3 | 1.0 |
| Vertical $\beta_y^*$ [mm] | 0.8 | 1.0 | 1.0 | 1.6 |
| Horizontal size at IP $\sigma_x^*$ [$\mu$m] | 6.4 | 13.0 | 13.7 | 38.2 |
| Vertical size at IP $\sigma_y^*$ [nm] | 28 | 41 | 36 | 68 |
| Energy spread (SR/BS) $\sigma_\delta$ [%] | 0.038/0.132 | 0.066/0.131 | 0.099/0.165 | 0.150/0.192 |
| Bunch length (SR/BS) $\sigma_z$ [mm] | 3.5/12.1 | 3.0/6.0 | 3.15/5.3 | 1.97/2.54 |
| Piwinski angle (SR/BS) | 8.2/28.5 | 3.5/7.0 | 3.4/5.8 | 0.8/1.0 |
| RF frequency [MHz] | 400 | 400 | 400 | 400 / 800 |
| RF voltage [GV] | 0.1 | 0.75 | 2.0 | 4.0 / 6.9 |
| Synchrotron tune $Q_s$ | 0.0250 | 0.0506 | 0.0358 | 0.0872 |
| Long. damping time [turns] | 1273 | 236 | 70.3 | 20.4 |
| Energy acceptance (DA) [%] | ±1.3 | ±1.3 | ±1.7 | −2.8, +2.4 |
| Luminosity / IP [$10^{34}$ cm$^{-2}$s$^{-1}$] | 230 | 28 | 8.5 | 1.55 |
| Beam-beam $\xi_x/\xi_y$ | 0.004/0.133 | 0.010/0.113 | 0.016/0.118 | 0.099/0.126 |
| Lifetime by rad. Bhabha [min] | 68 | 59 | 38 | 39 |
| Actual lifetime by BS [min] | > 200 | > 200 | 18 | 18 |

### 3.3 Luminosity Parameters and Run Plan

Table 5 presents the nominal luminosity, integrated luminosity per year, physics goals and resulting running time for the different modes of operation, based on the assumptions laid out above. This yields the timeline shown in Figure 1.

Phase 1 comprises two years of running-in and the full Z pole operation, W threshold scans and Higgs production modes. It can be accomplished within eight years. After one additional year of shutdown and staging of the RF, operation Phase 2, covering the top quark studies, would last for another five years. The entire FCC-ee physics programme could be completed within 14 to 15 years. After Phase 1 there could be a natural break point, where the decision might be taken, for example, not to upgrade to Phase 2, but to install the next hadron collider instead.

**Table 5:** Peak luminosity per IP, total luminosity per year (two IPs), luminosity target and run time for each FCC-ee working point.

| working point | luminosity $[10^{34} \text{ cm}^{-2}\text{s}^{-1}]$ | tot. lum./year $[\text{ab}^{-1}]$ / year | goal $[\text{ab}^{-1}]$ | run time [years] |
|---|---|---|---|---|
| Z first two years | 100 | 26 | 150 | 4 |
| Z other years | 200 | 52 | | |
| W | 32 | 8.3 | 10 | $\sim 1$ |
| H | 7.0 | 1.8 | 5 | 3 |
| RF reconfiguration | | | | 1 |
| t$\bar{\text{t}}$ 350 GeV (first year) | 0.8 | 0.20 | 0.2 | 1 |
| t$\bar{\text{t}}$ 365 GeV | 1.5 | 0.38 | 1.5 | 4 |

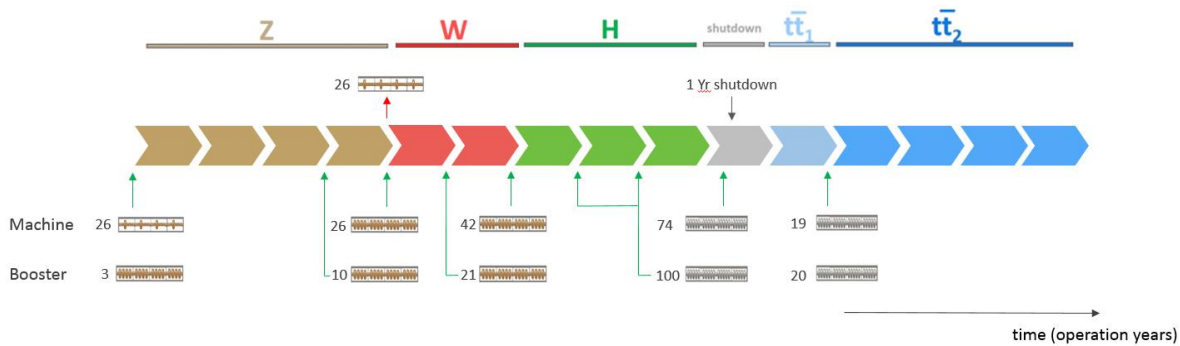

**Figure 1:** FCC-ee operation timeline. The bottom part indicates the number of cryomodules to be installed in the collider and booster respectively, during the various shutdown periods.

Conservatively, it is assumed that there will be another year at half the design luminosity after the one-year shutdown, for the first year of top running. This first year of the FCC-ee Phase-2 operation is performed at a beam energy of 175 GeV.

## 4    Linear Lepton Collider: CLIC

CLIC is a proposed future linear collider to provide e$^+$e$^-$ collisions up-to the multi-TeV energy range. The CLIC accelerator complex is based on normal-conducting high-frequency (X-band, 12 GHz) radiofrequency structures. It features e$^+$ and e$^-$ injectors, damping rings and the main linacs with their respective beam-delivery sections, as well as an electron drive beam complex for RF power generation and distribution. The construction of CLIC is planned in three stages [11], with the first stage at 380 GeV c.m. energy, an intermediate stage at 1.5 TeV and the final stage at 3 TeV. The key parameters for the accelerator are summarised in Table 6.

### 4.1    Estimating Annual Performance

The CLIC availability is based on the following model yearly operation model:

- 17 weeks for the extended winter shutdown (120 days)
- 30 days for commissioning
- 20 days for machine development
- 10 days for technical stops

This leaves 185 days of physics operation per year. An efficiency of 75% is assumed providing 1.2 10$^7$ seconds of yearly luminosity time. These assumptions are identical to those made for FCC-ee.

The assumed availability of CLIC is motivated by comparing to the performance of modern light-sources (FEL-linacs and low emittance rings). More information can be found in the CLIC project plan documentation being prepared for the European Strategy Update. The luminosity performance and improvements beyond the current baseline parameters are also discussed there.

A full year of commissioning of the CLIC complex is included as part of the construction period, in addition to commissioning periods foreseen for the individual systems. A luminosity ramp-up over three years at 10, 30, 60% of the final luminosity for stage 1, and over two years at 25, 75% for stages 2 and 3, is foreseen. The scenario above is the same as used for the ILC, with the exception that the ILC luminosity year is foreseen to be $1.6 \, 10^7$ sec. The resulting run plan in terms of integrated luminosity, tuned to the levels specified by the CLIC physics study group for each stage, is shown in Figure 2. The run programme is based on making decisions about a potential next stage after 3-5 years of operation at a given stage, guided by physics results at that time. This allows construction of a next stage partly in parallel with continued physics operation.

**Table 6:** Key parameters of the CLIC linear e+e- collider.

| Parameter | Symbol [Unit] | CLIC | | |
|---|---|---|---|---|
| C.M. Energy | $E_{CM}$ [GeV] | 380 | 1500 | 3000 |
| Gradient | $G$ [MV/m] | 72 | 72 & 100 | 72 & 100 |
| Length | [km] | 11.4 | 29.0 | 50.1 |
| Repetition rate | $f_{rep}$ [Hz] | 50 | 50 | 50 |
| Bunches per train | $n$ | 352 | 312 | 312 |
| Particles per bunch | $N$ [$10^9$] | 5.2 | 3.7 | 3.7 |
| Rms bunch length | $\sigma_z$ [$\mu m$] | 70 | 44 | 44 |
| Rms energy spread | [%] | 0.35 | 0.35 | 0.35 |
| Normalized rms emittances | $\gamma \varepsilon_{x,y}$ [nm] | 950, 30 | 660, 20 | 660, 20 |
| Rms IP beam size | $\sigma_{x,y}$ [nm/nm] | 149, 3 | 60, 1.5 | 40, 1 |
| IP beta function | $\beta^*_{x,y}$ [mm] | 8, 0.1 | 8, 0.1 | 6, 0.07 |
| Luminosity | $L$ [$10^{34}$ cm$^{-2}$s$^{-1}$] | 1.5 | 3.7 | 6.0 |
| Luminosity within 1% of $\sqrt{s}$ | $L_{0.01}$ [$10^{34}$ cm$^{-2}$s$^{-1}$] | 0.9 | 1.4 | 2.0 |
| Int. luminosity per year | $L_{int}$ [fb$^{-1}$] | 180 | 444 | 720 |

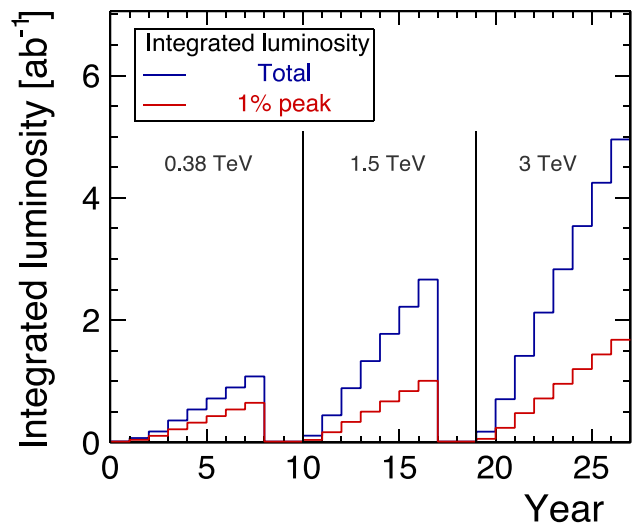

**Figure 2:** Projected CLIC integrated luminosity for stages 1 to 3 [11].

## 5    Lepton-Hadron Colliders: (HE-)LHeC and FCC-eh

### 5.1    Concept and Key Parameters for Lepton-Proton Collisions

Lepton-proton collisions can be achieved by colliding protons circulating in the HL-LHC, HE-LHC or FCC-hh with a 60 GeV polarised electron beam. For the (HL-)LHC, this option is known as the Large Hadron electron Collider (LHeC) [12, 13]. The electron beam would be provided by a dedicated 3-pass recirculating energy-recovery linac (ERL). Since this machine is not housed in the (HE-)LHC or FCC tunnel, it minimises any interference with the main hadron beam infrastructure. To a large extent, the electron accelerator may be built independently from the operation of the proton machine. The same ERL machine could be used for the LHeC, for the HE-LHeC and later on for the FCC-eh, presenting a potential large return on the initial investment.

In a simplified model, the luminosity $L$ of the HE-LHeC is given by [12, 13]:

$$L = \frac{N_p N_e f}{4\pi \varepsilon_p \beta_p} \cdot H_{\text{geom}} H_{b-b} H_{\text{coll}}$$

where $N_p$ is the number of protons per bunch and $\varepsilon_p$ and $\beta_p$ are the geometric proton beam emittance and proton IP beta function. It is assumed that the proton-beam parameters $N_p$ and $\varepsilon_p$ are governed by the main experiments that collide protons on protons, as the baseline is for concurrent e-p and p-p operation. The bunch frequency, denoted by $f = 1/\Delta$, is 40 MHz for the default bunch spacing of $\Delta = 25$ ns. $N_e$ is the number of electrons per bunch, which determines the electron current $I_e = eN_e f$. The factors $H_{geom}$, $H_{b-b}$ and $H_{coll}$ are geometric correction factors with values typically close to unity. $H_{\text{geom}}$ is the reduction of the luminosity due to the hourglass effect, $H_{b-b}$ is the increase of the luminosity by the strong attractive beam-beam forces and $H_{\text{coll}}$ is a factor that takes the filling pattern of the electron and the proton beam into account. Estimates for these parameters are shown in Table 7.

Compared to the CDR of the LHeC from 2012 (where a luminosity of $10^{33}$ cm$^{-2}$s$^{-1}$ was suggested), it seems possible to achieve peak luminosities near to or larger than $10^{34}$ cm$^{-2}$s$^{-1}$, which makes these future ep colliders efficient machines for the study of new physics at the accelerator energy frontier.

**Table 7:** Baseline parameters and estimated peak luminosities of future electron-proton collider configurations based on an electron ERL [14].

| Parameter [unit] | LHeC CDR | ep at HL-LHC | ep at HE-LHC | FCC-he |
|---|---|---|---|---|
| $E_p$ [TeV] | 7 | 7 | 13.5 | 50 |
| $E_e$ [GeV] | 60 | 60 | 60 | 60 |
| $\sqrt{s}$ [TeV] | 1.3 | 1.3 | 1.7 | 3.5 |
| Bunch spacing [ns] | 25 | 25 | 25 | 25 |
| Protons per bunch [$10^{11}$] | 1.7 | 2.2 | 2.5 | 1 |
| $\gamma\varepsilon_p$ [$\mu$m] | 3.7 | 2 | 2.5 | 2.2 |
| Electrons per bunch [$10^9$] | 1 | 2.3 | 3.0 | 3.0 |
| Electron current [mA] | 6.4 | 15 | 20 | 20 |
| IP beta function  $\beta_p^*$ [cm] | 10 | 7 | 10 | 15 |
| Hourglass factor  $H_{\text{geom}}$ | 0.9 | 0.9 | 0.9 | 0.9 |
| Pinch factor  $H_{b-b}$ | 1.3 | 1.3 | 1.3 | 1.3 |
| Proton filling  $H_{\text{coll}}$ | 0.8 | 0.8 | 0.8 | 0.8 |
| Luminosity [$10^{33}$ cm$^{-2}$s$^{-1}$] | 1 | 8 | 12 | 15 |

## 5.2 Run Plan and Expected Performance

Assumptions and expected luminosity performance for three LHeC data-taking periods are compiled in Table 8. The projected cumulative luminosity evolution of LHeC is illustrated in Figure 3.

Three running modes are distinguished:

1. LHeC during LHC Run 5: initial operation concurrent to pp, yielding 50 fb$^{-1}$. The peak luminosity is 100 times higher than for HERA, and collisions occur at higher energies. This run will address SM precision physics, PDFs, etc.
2. LHeC during LHC Run 6: design operation concurrent to pp, adding another 175 fb$^{-1}$
3. A final LHeC run in dedicated operation without pp adds a further 650 fb$^{-1}$, and brings the total integrated luminosity close to 1 ab$^{-1}$. This is the era of high-precision Higgs physics and rare processes.

Other short runs (a few fb$^{-1}$) at low electron energy and three months for eA are not yet scheduled. In addition, runs at lower proton energy could be of interest. For each period, it is assumed that in year 1, the machine will operate at only half of the peak luminosity.

**Table 8:** Parameters and expected performance for the LHeC data-taking periods.

| Parameter | Unit | Initial: Run 5 | Design: Run 6 | Dedicated |
|---|---|---|---|---|
| Brightness $N_p/(\gamma\epsilon_p)$ | $10^{17}$ m$^{-1}$ | 2.2/2.5 | 2.2/2.5 | 2.2/2.5 |
| electron beam current $I_e$ | mA | 15 | 5 | 50 |
| proton $\beta^*$ | m | 0.1 | 0.07 | 0.07 |
| peak luminosity | $10^{34}$ cm$^{-2}$s$^{-1}$ | 0.5 | 1.2 | 2.4 |
| p beam lifetime | h | 16.7 | 16.7 | 100 |
| fill duration | h | 11.7 | 11.7 | 21 |
| turnaround time | h | 4 | 4 | 3 |
| overall efficiency | % | 54 | 54 | 60 |
| Physics time / year | days | 160 | 180 | 185 |
| Annual integrated lumin. | fb$^{-1}$ | 20 | 50 | 180 |

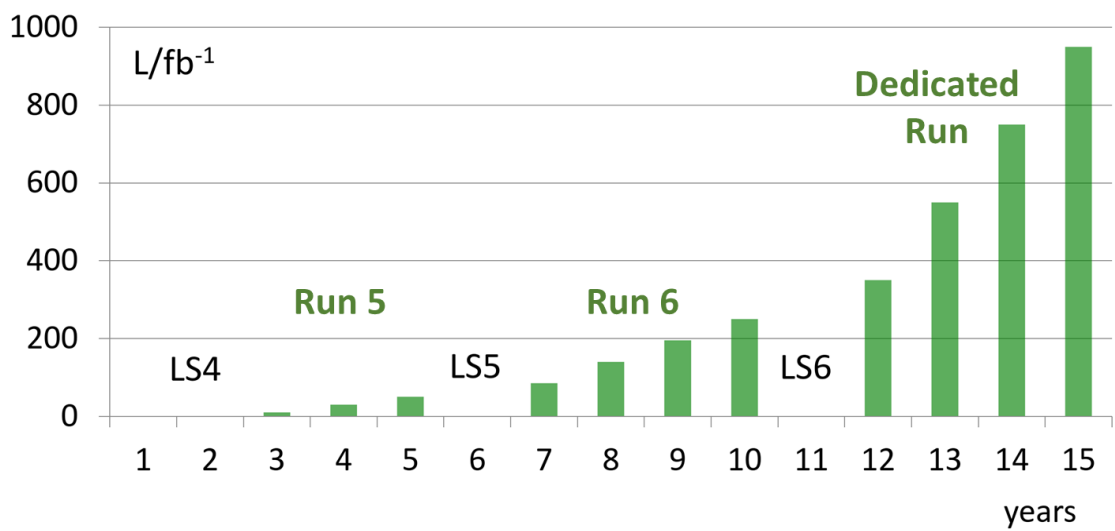

**Figure 3:** Projected LHeC cumulative integrated luminosity.

LHeC would have seven years of concurrent ep and pp operation, if LHeC begins after LS4 and if LHC Run 6 lasts four years, as it is desirable to maximise the overlap of ep with pp physics. Three data-taking periods are considered to include a final 'bridge' time to end LHC. The last HL-LHC run and the final bridge period are assumed to last four years each.

HE-LHeC and FCC-eh could provide even higher luminosities at even higher energies. The running time is 20 years instead of seven. It can be assumed that e-p is there from the start. A total luminosity of the order of 2 ab$^{-1}$ can therefore be reached, even when running only in parasitic mode. FCC-eh could deliver 1.2 ab$^{-1}$ in 10 years for 20 mA electron current. Following an initial phase, regular operation will be determined by the pp colliders.

## 5.3 Ion-Lepton Collisions

The heavy-ion beams that the CERN injector complex can provide are also a unique basis for high-energy, high-luminosity deep inelastic electron-ion scattering physics. Combining the intense beams of $^{208}$Pb$^{82+}$ nuclei that are to be provided for HL-LHC, HE-LHC, FCC-hh with the default 60 GeV electron ERL yields the eA parameter set of Table 9.

Radiation damping of Pb beams in the hadron rings is about twice as fast as for protons and can be fully exploited. For the HE-LHC and FCC-hh cases, the emittance values in Table 9 are estimates of the effective average values during a fill in which Pb-Pb collisions are being provided at one other interaction point [15].

**Table 9:** Baseline parameters of future electron-ion collider configurations based on the electron ERL, in concurrent eA and AA operation mode [14].

| parameter [unit] | LHeC (HL-LHC) | eA at HE-LHC | FCC-he |
|---|---|---|---|
| $E_{Pb}$ [PeV] | 0.574 | 1.03 | 4.1 |
| $E_e$ [GeV] | 60 | 60 | 60 |
| $\sqrt{s_{eN}}$ electron-nucleon [TeV] | 0.8 | 1.1 | 2.2 |
| Bunch spacing [ns] | 50 | 50 | 100 |
| No. of bunches | 1200 | 1200 | 2072 |
| Ions per bunch [$10^8$] | 1.8 | 1.8 | 1.8 |
| $\gamma\epsilon_A$ [$\mu$m] | 1.5 | 1.0 | 0.9 |
| Electrons per bunch [$10^9$] | 4.67 | 6.2 | 12.5 |
| Electron current [mA] | 15 | 20 | 20 |
| IP beta function $\beta_A^*$ [cm] | 7 | 10 | 15 |
| Hourglass factor H$_{geom}$ | 0.9 | 0.9 | 0.9 |
| Pinch factor H$_{b-b}$ | 1.3 | 1.3 | 1.3 |
| Bunch filling H$_{coll}$ | 0.8 | 0.8 | 0.8 |
| Luminosity [$10^{32}$ cm$^{-2}$s$^{-1}$] | 7 | 18 | 54 |

## 6 Acknowledgement

## 7 References

[1] G. Apollinari, I. Bejar Alonso, O. Bruning, P. Fessia, M. Lamont, L. Rossi and L. Tavian, "High-Luminosity Large Hadron Collider (HL-LHC) : Technical Design Report V. 0.1," CERN-2017-007-M, Geneva, 2017.

[2] O. Bruning, P. Collier, P. Lebrun, S. Myers, R. Ostojic, J. Poole and P. P., "LHC Design Report , v.1 : the LHC Main Ring," CERN-2004-003-V-1, Geneva, 2014.

[3] J. Jowett, "Colliding Heavy Ions in the LHC," 9th International Particle Accelerator Conference (IPAC 2018), Vancouver, BC, Canada, 2018.

[4] M. Schaumann and others, "First Xenon-Xenon Collisions in the LHC," 9th International Particle Accelerator Conference (IPAC 2018), Vancouver, BC, Canada, 2018.

[5] M. Benedikt, D. Schulte and F. Zimmermann, "Optimizing integrated luminosity of future hadron colliders," *Phys. Rev. ST Accel. Beams* , vol. 18, p. 101002, 2015.

[6] M. Benedikt and F. Zimmermann, "Proton Colliders at the Energy Frontier," *Nucl. Instr. Methods A,* no. Special Kai Siegbahn Issue, 2018.

[7] F. Zimmermann and others, "High-Energy LHC Design," *9th International Particle Accelerator Conference (IPAC 2018), Vancouver, BC Canada, 2018,* 2018.

[8] H. Damerau, A. Funken, R. Garoby, S. Gilardoni, B. Goddard, K. Hanke, A. Lombardi, D. Manglunki, M. Meddahi, B. Mikulec, G. Rumolo, E. Shaposhnikova and M. C. J. Vretenar, "LHC Injectors Upgrade, Technical Design Report, Vol. I: Protons," CERN-ACC-2014-0337, Geneva, 2014.

[9] M. Bicer and others, "First look at the physics case of TLEP," *Journal of High Energy Physics,* p. 164, 2014.

[10] P. Janot, "Perspectives for Future Circular Colliders (1/3)," CERN Academic Training, 11 October 2017, CERN, Geneva. Switzerland, 2017.

[11] M. Boland and others, "Updated baseline for a staged Compact Linear Collider," CERN-2016-004 and arXiv:1608.07537, 2016.

[12] J. Abelleira Fernandez and others, "A Large Hadron Electron Collider at CERN," *Journal of Physics G: Nuclear and Particle Physics,* vol. 7, p. 075001, 2012.

[13] F. Zimmermann, O. Bruning and M. Klein, "The LHeC as a Higgs Boson Factory," Proc. IPAC2013, Shanghai, China , 2013.

[14] O. Bruning, J. Jowett, M. Klein, D. Pellegrini, D. Schulte and F. Zimmermann, "Future Circular Collider Study FCC-eh Baseline Parameters," CERN FCC-ACC-RPT-012, 2017.

[15] A. Dainese and others, "Heavy ions at the Future Circular Collider," arXiv:1605.01389, 2016.